\documentclass[traditabstract]{aa}  

\usepackage{natbib}
\usepackage{hyperref}
\bibpunct{(}{)}{;}{a}{}{,} 
\usepackage{graphicx}
\usepackage{txfonts}
\begin{document} 

\newcommand {\CaII} {\ion{Ca}{II}}
\newcommand{\ltau} {$\log \tau_{500}$}
   \title{Mapping the Sun's upper photosphere with artificial neural networks}

   \author{H. Socas-Navarro
          \inst{1,2}
          \and
          A. Asensio Ramos\inst{1,2}
          }

   \institute{Instituto de Astrof\'\i sica de Canarias, V\'\i a L\'actea S/N, La Laguna 38205, Tenerife, Spain\\
   \email{hsocas@iac.es, andres.asensio@iac.es}
         \and
             Departamento de Astrof\'\i sica, Universidad de La Laguna, 38205, Tenerife, Spain \\
             }

   \date{Received ; accepted }

 
  \abstract
   {We have developed an inversion procedure designed for high-resolution solar spectro-polarimeters, such as Hinode/SP or DKIST/ViSP. The procedure is based on artificial neural networks trained with profiles generated from random atmospheric stratifications for a high generalization capability. When applied to Hinode data we find a hot fine-scale network structure whose morphology changes with height. In the middle layers this network resembles what is observed in G-band filtergrams but it is not identical. Surprisingly, the temperature enhancements in the middle and upper photosphere have a reversed pattern. Hot pixels in the middle photosphere, possibly associated to small-scale magnetic elements, appear cool at the \ltau=$-3$ and $-4$ level, and viceversa. Finally, we find hot arcs on the limb side of magnetic pores, which we interpret as the first direct observational evidence of the "hot wall" effect in temperature.}
   
   {}


   \maketitle
%

\section{Introduction}

Inversion techniques allow us to retrieve information encoded in spectral lines about the
atmospheres where they form. A wide variety of strategies have been employed for decades
in solar physics to interpret spectroscopic and spectropolarimetric observations (see e.g.
the reviews by \citealt{dTIRC16,BR06,SN01}). Most applications are based on the
least-squares fitting of the observed spectral lines with synthetic profiles, which are
computed from model atmospheres whose parameters are iteratively adjusted until a
satisfactory fit is attained. However, advances in instrumentation are driving an
increasing interest in the exploration of alternative methods. Two-dimensional
spectropolarimetry is now very common and fast growing data rates motivate the exploration
of new algorithms that have the potential of being faster and/or more robust for
systematic application.

Artificial neural networks (ANNs) offer a promising new approach for many purposes where
profile fitting is inadequate because one needs a faster or a more robust performance. The
first applications of ANNs in solar physics are almost 20 years old, dating back to
\citet{CS01} and \citet{SN02}. However, while those first efforts produced encouraging
results, ANN inversions were not immediately adopted by the community, for two reasons
mainly. First, disentangling the magnetic filling factor from the intrinsic field strength
has proven extremely challenging, as noted since those early works \citep{SN03}. The
magnetic field tends to exhibit small-scale structures in the solar photosphere. In
arc-second resolution observations, it is common to find pixels where the magnetic field
occupies less than 10\% of the resolution element. This area fraction is referred to
as the filling factor and it introduces an important complication for ANN inversions.
Second, a more practical issue is the complexity involved in the coding of algorithms for
the training of an ANN model. 

Those early problems have been largely resolved in recent years, leading to a renewed
interest in ANNs   \citep[e.g.,][]{LXW+20,GBD+20,DBAA18, AADB19,FAA19,MG20}. Current (and
upcoming) instrumentation are delivering very high resolution observations, mitigating the
filling factor problem. Furthermore, there has been a tremendous development in the field
of deep learning and many sophisticated tools have been made publicly available to
simplify the problem of building and training ANNs \citep[e.g.,][]{tensorflow2015-whitepaper,NEURIPS2019_9015}.

In this paper we use a relatively simple ANN model and take a different approach to
previous work for the training strategy. Instead of using a simulation snapshot as the
starting point for a training set, as in \cite{AADB19} or \cite{MG20}, we create a
database of profiles from random stratifications of the relevant parameters. This provides
a wide coverage of the parameter space and guarantees that the ANN is not specialized on
any particular scenario. Unlike \cite{AADB19}, whose ANN performs a full inversion of the
entire 2D field at once, this ANN works on each pixel independently. In that regard, it is
more similar to a traditional inversion technique. 

We created two different ANNs, one to invert photospheric observations from DKIST/ViSP
(Daniel K Inouye Solar Telescope/Visible Spectro-Polarimeter, see \citealt{RWK+20}) and
the other one for the Hinode satellite's SOT/SP (Solar Optical
Telescope/Spectro-Polarimeter). DKIST/ViSP data are not yet available so we focus here on
the analysis of the Hinode inversions. After testing the procedure with synthetic data and
previous inversions of real observations, we applied it to Hinode observations of active
regions. In this manner we obtained datacubes with a fairly unique combination of high
spatial resolution, large field of view and depth-dependent temperatures. These maps show
a fine hot network in active regions, particularly around sunspots and pores. 

We find some surprising results in this application, such as an anticorrelation between hot pixels in middle and upper layers. Also, the inversions reveal a series of hot arcs running along the limb
side of pores in the observed regions. We interpret these arcs as the first direct observation of the "hot wall" effect, a prediction of fluxtube models since the work of \cite{S76} which had not been directly observed thus far. 

\section{The ANN model and training set}
\label{training}

All the calculations presented in this paper were produced with relatively standard
computer hardware. We employed a Linux workstation powered by eight 3·GHz Intel Xeon cores.
The system is equipped with a GTX 1080 GPU that handles most of the
ANN-related processing. Our ANN model and codes are publicly available in a repository{\footnote{\href{https://github.com/hsocasnavarro/Paper_SNAR21}{https://github.com/hsocasnavarro/Paper\_SNAR21}}}.

The ANN is created and trained using PyTorch \citep{NEURIPS2019_9015}. It is a simple multilayer
perceptron with six hidden layers between the input and output layers.
Each hidden layer comprises 300 neurons. The input layer has a number of neurons that
matches the number of spectral pixels in a given profile (175 for DKIST/ViSP and 112 for
Hinode SOT/SP). The output layer has 9 neurons, which correspond to the output parameters
that we wish to retrieve. These parameters are: 5 temperatures at different heights, 3
components of the magnetic field vector and a single-valued line-of-sight velocity. In all
cases, the activation function chosen is a leaky ReLU \citep{Maas13rectifiernonlinearities}. 
The entire training procedure takes a few hours on our hardware described
above.

For the training and validation sets we compute one million synthetic profiles from randomized model
atmospheres. A thousand models and profiles are taken as the validation set and the rest are used for training. These models are obtained as random variations from four different reference
atmospheres, namely: HSRA \citep{GNK+71}, VAL-C \citep{VAL81}, FAL-C \citep{FAL93} and the
sunspot model M of \citet{MAC+86}. We provide here a description of the randomization
procedure in some detail because the construction of this database is critical for the
ability of the ANN to perform adequately when faced with real observations and to exhibit
good generalization properties.

For each relevant parameter, we take the stratification in the reference atmosphere and
add a depth-dependent perturbation to it. The perturbation is constructed by assigning
values to certain layers and then interpolating in depth. In the case of temperature, the
parameter to which spectral lines are most sensitive, we start by creating a perturbation
at four layers. These layers are not necessarily the same heights that the ANN will
retrieve. They are equispaced in the logarithm of continuum optical depth at 500~nm
(\ltau) and their actual location is different for the four reference
atmospheres. The perturbations at these four points are drawn from a Gaussian distribution
with a 1,500~K standard deviation. From these four values, the depth-dependent perturbation is
interpolated to the entire grid and added to the reference model.
With the new thermal stratification, the model is set in hydrostatic equilibrium and the
equation of state is solved to compute plasma densities, ionization fractions, relevant
molecules and electron densities.

For the magnetic field, we have that $B_z$ (the line-of-sight component) is linear in
\ltau, whereas $B_x$ and $B_y$ (the transverse components) are constant with height.
$B_z$ is defined by $B_z(0)$, its value at \ltau~$=0$, and its gradient. We construct three
possible scenarios with weak, strong and extreme fields, having probabilities of 45\%,
45\% and 10\% respectively. The field strength $B_z(0)$ takes values from a uniform
distribution with a width of 500, 2000 and 6000~$G$ for the weak, strong and extreme
fields, respectively. The sign for each field component is randomly set to $\pm 1$
except for $B_x$, which is always taken as positive. Since the Zeeman effect has a
180-degree ambiguity in the transverse component of the field, we restrict our solutions
to the subspace with positive $B_x$. The $B_z$ gradient is set to either 0 or a random
value, with a 50\% probability. The random value is taken from a uniform probability
distribution between -150 and 150~$G$ per unit in \ltau .

The filling factor ($\alpha$) is set to 1 in 50\% of the models. The rest have a uniform
distribution between 0.1 and 1. In addition to the filling factor, we consider a fixed
amount of stray light in the instrument by adding an average quiet Sun profile to
Stokes~I. The amount of stray light is fixed to 10\%, which is a typical value for
spectrographs.

This training set was built with the purpose of covering a sufficiently wide range of
profiles for the ANN to work with all possible observations of the solar photosphere in
the 630~nm spectral window observed by Hinode. The statistical distribution of our random
atmospheres is not necessarily optimal. We relied on past experience and numerical
experimentation to determine a suitable set. A systematic analysis is beyond the scope of
this work.

We used NICOLE \citep{SNdCRAR+15} to compute synthetic Stokes profiles for the entire set
of one million random models in the database. The synthesis parameters for one of the
training sets were defined to mimic Hinode/SP observations.  DKIST/ViSP will also feature
a preset mode to observe the same 630~nm window so we produced another similar set of
profiles simulating those observations in anticipation of its science operations. Both
training sets are publicly available in the repository mentioned above.

\begin{figure*}
    \centering
    \includegraphics[height=0.9\textheight]{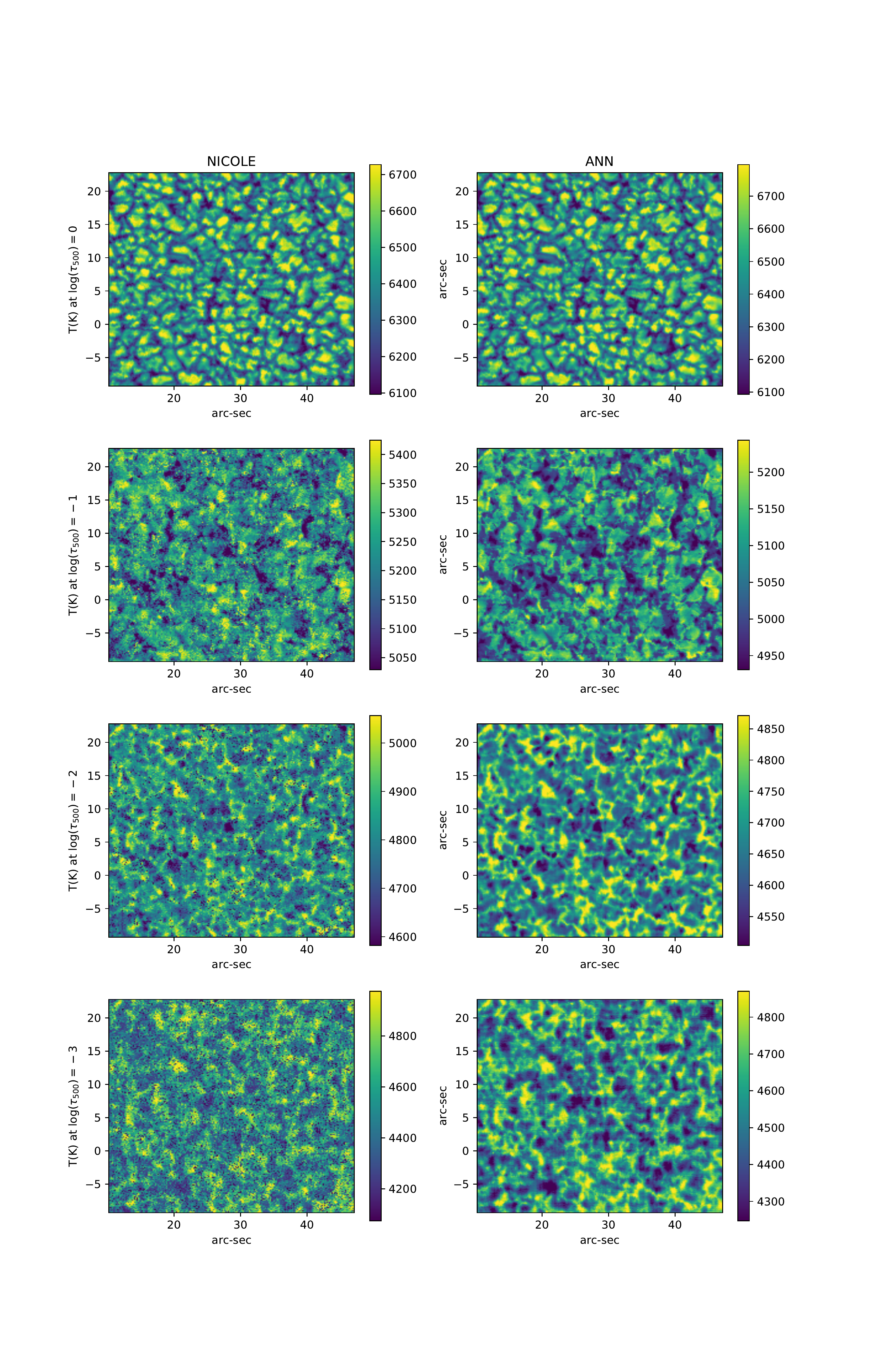}
    \caption{Comparison of inversions of a Hinode map performed with NICOLE (left) and our ANN (right).}
    \label{quiet1}
\end{figure*}

The Stokes profiles are fed as inputs to the ANN. For the outputs, we extract a set of 9
parameters from the random model atmospheres in the database. These parameters are: five
temperatures ($T_0$, ..., $T_4$), extracted at optical depths \ltau = $0$,...,$-4$, a bulk
Doppler velocity ($v_z$) and the three components of the pixel-averaged magnetic field
($F_x$, $F_y$, $F_z$, where $F_i=\alpha B_i$ for $i=x,y,z$). We do not aim here at
disentangling the filling factor $\alpha$ from the intrinsic magnetic field strength
($B_i$) in the magnetic element. We seek to retrieve the magnetic flux density ($F$) in
the resolution element, which simplifies the problem. 

\begin{figure*}
    \centering
    \includegraphics[height=.8\textheight]{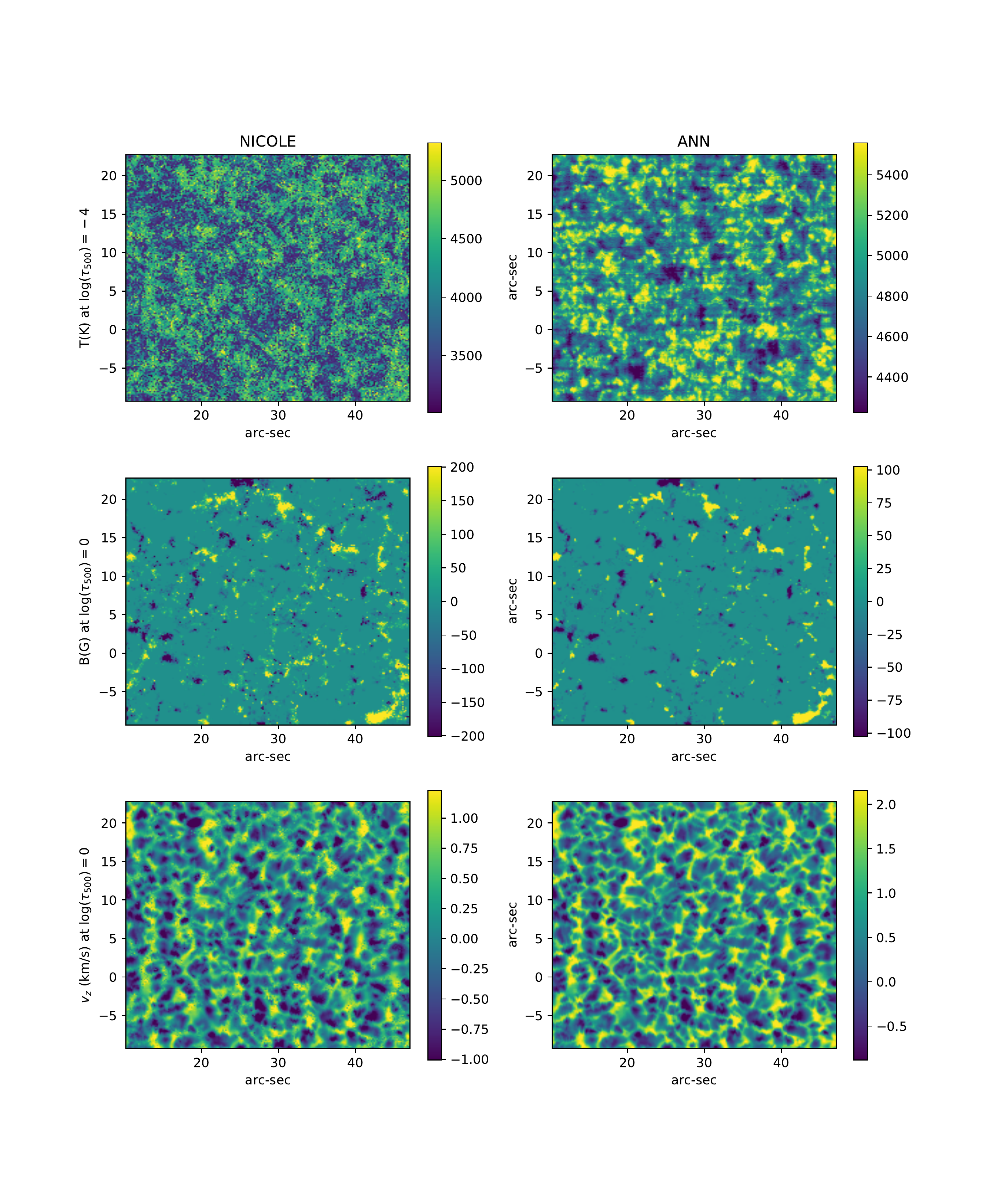}
    \caption{Comparison of inversions of a Hinode map performed with NICOLE (left) and our ANN (right). Positive (negative) velocities are directed downards (upwards). Positive (negative) magnetic polarity represents fields pointing up (down) from the solar surface.}
    \label{quiet2}
\end{figure*}

\section{Comparisons with other inversions}
\label{tests}

\begin{figure*}
    \centering
    \includegraphics[height=.8\textheight]{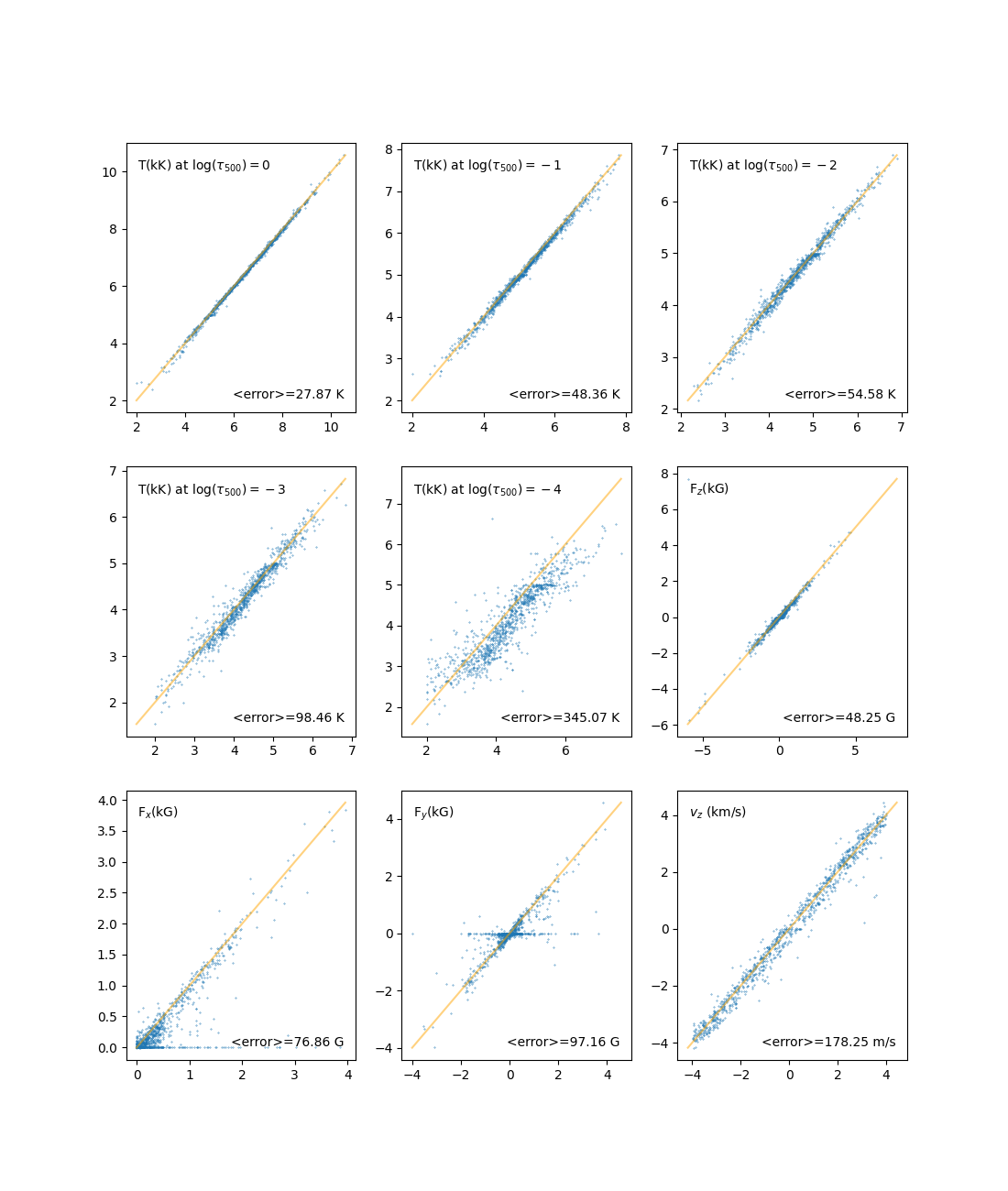}
    \caption{Tests of the ANN performance with the validation set. Abscissas are the "true" values and ordinates are the ANN outputs. Each plot shows the average (median) error in the recovery of that parameter.}
    \label{fig:validation}
\end{figure*}

After successfully training the ANN and observing a good recovery of the validation set (see Figure~\ref{fig:validation}), we tested it with real observations. As noted in
previous work \citep{SN05e}, a good performance with the validation set composed of
synthetic observations does not guarantee a good operation with real data. 

Ideally, one would like to have inversions of Hinode/SP data to compare with our ANN.
Unfortunately, there are very few inversions of Hinode/SP maps that yield the height
stratification. The standard pipeline includes an inversion carried out by the instrument
team with the MERLIN code \citep{LCG+07}, which is based on the Milne-Eddington
approximation and therefore does not provide information on the height dependence of any
physical quantities. One of the few inversions with the height stratification existing in
the literature is that of \citet{SN11} using NICOLE (later refined in \citealt{SN15}). 

We took the same Hinode/SP observations used for the NICOLE inversions and processed them
with our ANN. The NICOLE inversions took about 5 hours on a dedicated parallel run over
the eight cores of our workstation. The ANN inversion was completed in half a second.

\begin{figure*}
    \centering
    \includegraphics[width=\textwidth]{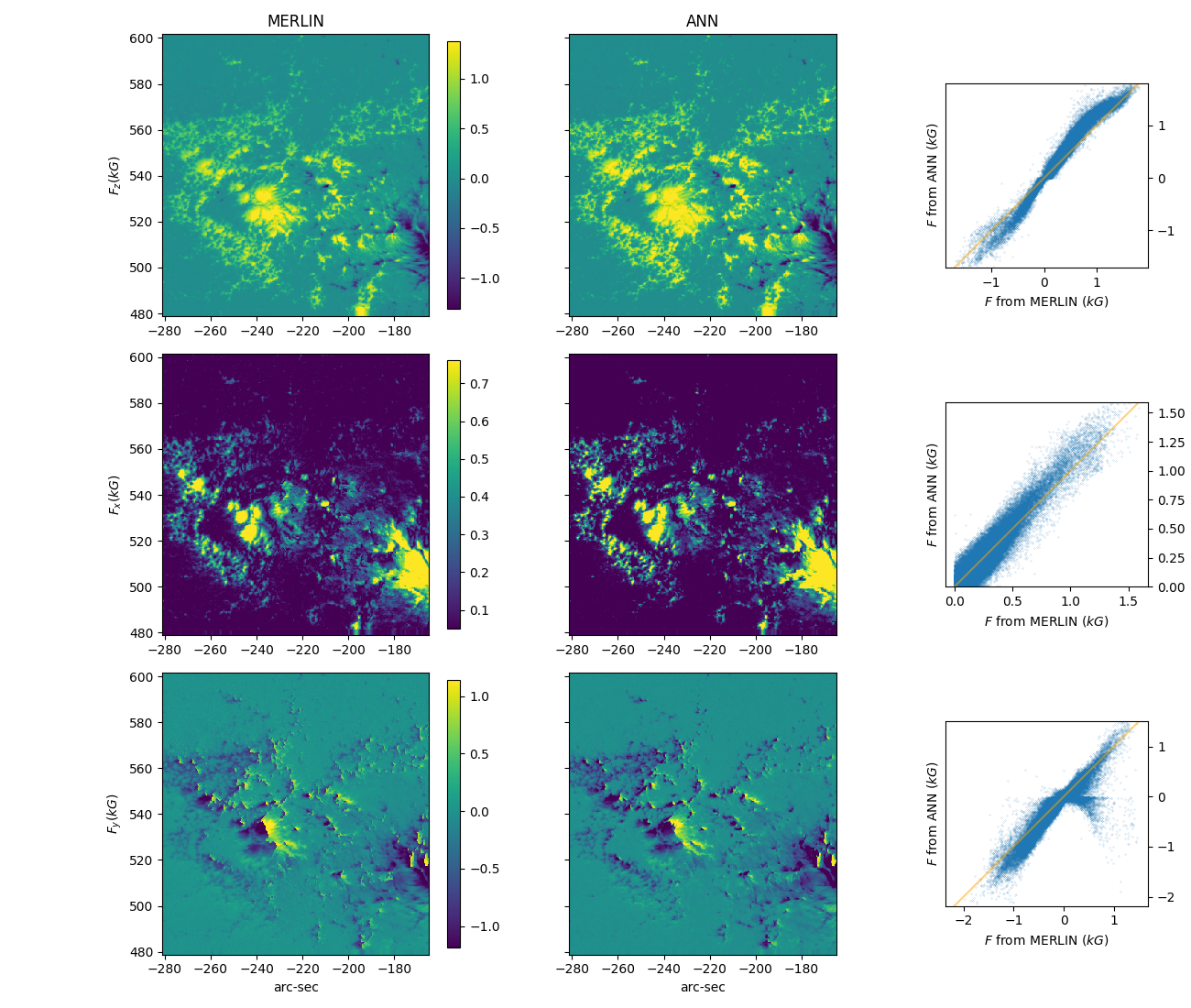}
    \caption{Comparison of inversions of a Hinode map (map1) performed with MERLIN (left column) and our ANN (center). Scatter plots are shown in the right column. Positive (negative) magnetic polarity represents fields pointing up (down) from the solar surface.}
    \label{fig:ar1b}
\end{figure*}

\begin{figure*}
    \centering
    \includegraphics[width=\textwidth]{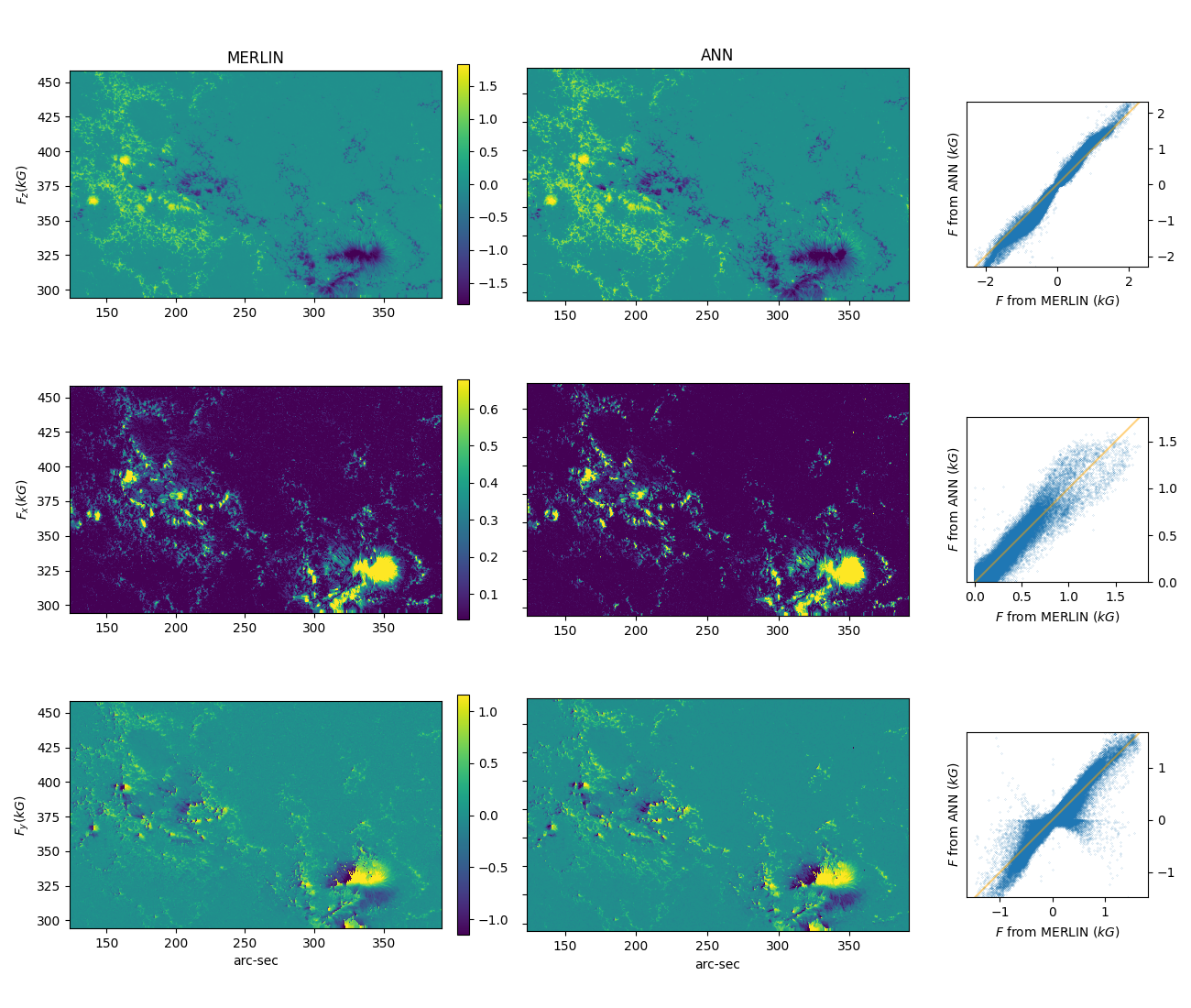}
    \caption{Comparison of inversions of a Hinode map (map2) performed with MERLIN (left column) and our ANN (center). Scatter plots are shown in the right column. Positive (negative) magnetic polarity represents fields pointing up (down) from the solar surface.}
    \label{fig:ar2b}
\end{figure*}

An additional postprocessing renormalization was applied on each ANN temperature output,
so that the average value would match that of the NICOLE inversions. This is done to
remove (at least to first order) some small residuals that arise in the application to real data. 

We do not find these residuals in the validation tests so they must be
due to systematic differences between our
synthetic training set and the real observations, such as observation artifacts, differences in the PSF or an inaccurate estimate of the
stray light used in the synthesis. A detailed analysis of these residuals is beyond the scope of this paper but for our purposes here, this simple renormalization (the same for all observations) resolves the issue.

The normalization factors for $T_0$,...,$T_4$ are 0.90,1.26,1.29,1.44 and
1.60, respectively. The growing trend of these factors indicates that the ANN produces
models that are, on average, steeper than those obtained with NICOLE. The synthesis tests
presented below demonstrate that the model atmospheres obtained in this manner produce
spectral profiles very similar to the observations. For the magnetic field (see below),
this calibration yields a factor of 0.7 in all three components. We incorporate this
normalization factor in all subsequent inversions.

A comparison of the maps produced by the ANN and those from NICOLE (the 2015 version) is
presented in Figs~\ref{quiet1} and~\ref{quiet2}. The similarity between the spatial
structures in the images obtained with both techniques is remarkable. The NICOLE
inversions are much more noisy, especially in the higher layers. ANNs are known to have
good noise filtering properties. In this case, most of the noise in the NICOLE data is
"inversion noise" produced by the specific $\chi^2$ fitting procedure that seeks the
best fit to the entire line profile. 
The upper layers are probed only by the core of the spectral lines. Since the core
occupies very few pixels in the spectral profile, there is very little information about
those upper layers. For very similar profiles, the $\chi^2$ minimization might reach
slightly different solution where the core is fitted with more or less accuracy, perhaps
compensating it with a better fit to other spectral regions. The end result is a
pixel-to-pixel variation that becomes more important in those layers where the profile is
less sensitive. This problem could be mitigated by  fine-tuning the weights, giving more
weight to the pixels that carry the relevant information. However, different layers would
require a different optimization.

The ANN, on the other hand, "learns" what are the optimal spectral points that it needs to
focus on for each layer. There is a direct, deterministic mapping between the observations
and the inversion result. For that reason the ANN maps (right column in the figures) look
cleaner. We can even see some of the residual defects in the data reduction that are still
present in the observations, as they propagate directly into the results.

The similarity between both sets of images confirms that NICOLE and the ANN are giving
consistent results. This test should not be viewed as NICOLE giving the "correct" answer
and our ANN being an approximation. Both techniques are approximations and the difference
between them is the sum of their respective errors.

The accuracy of the ANN in recovering the magnetic field is not relevant for the purposes
of this paper. Nevertheless, we show here similar comparisons for the sake of
completeness. We processed two active region maps observed with Hinode/SP (more details in
section~\ref{sec:ar} below). These maps are 384$\times$384 (map1) and 871$\times$512
(map2) spatial pixels. The ANN inversions took 4 and 11 seconds, respectively. Standard
inversions with the MERLIN code are available for these maps. Figures~\ref{fig:ar1b}
and~\ref{fig:ar2b} show the comparison of the magnetic flux inferred by MERLIN and our
ANN. For consistency with the training set, the 180-degree ambiguity is resolved by
choosing the solution that has a positive component along the $x$ axis.

\subsection{Reconstruction fits}

A common problem with ANN-based inversions is that they
are not based on fitting the observations, unlike $\chi^2$ fitting methods. 
The quality of a fit is usually a good indicator to assess the
validity of the results. Our approach suffers from this limitation too but it does provide
enough information to reconstruct a model atmosphere and, from there, synthesize spectral
profiles that can then be compared to the observations. It does not provide the same
information as a fit because the reconstruction of the atmosphere implies additional
approximations. Nevertheless, it is still useful information.

We take the parameters from the ANN inversion and compute model atmospheres by
interpolating them in optical depth. For the temperature stratification, we perform a
cubic interpolation of the five temperatures between \ltau~$=0$ and $-4$. Above \ltau$=-4$ we
impose that the stratification becomes flat. In the deeper layers below \ltau~$=0$, the
temperature gradient usually becomes steeper. After some experimentation, we concluded
that a gradient that is 30\% larger than between \ltau~$=0$ and $-1$ works best in reproducing
the observations. Hydrostatic equilibrium is imposed and the plasma equation of state is
solved numerically to determine gas and electron densities, ionization stages and relevant
molecules. The magnetic field and bulk Doppler velocities are taken as constant with
height from the ANN inversion. 

\begin{figure}
    \centering
    \includegraphics[width=.5\textwidth]{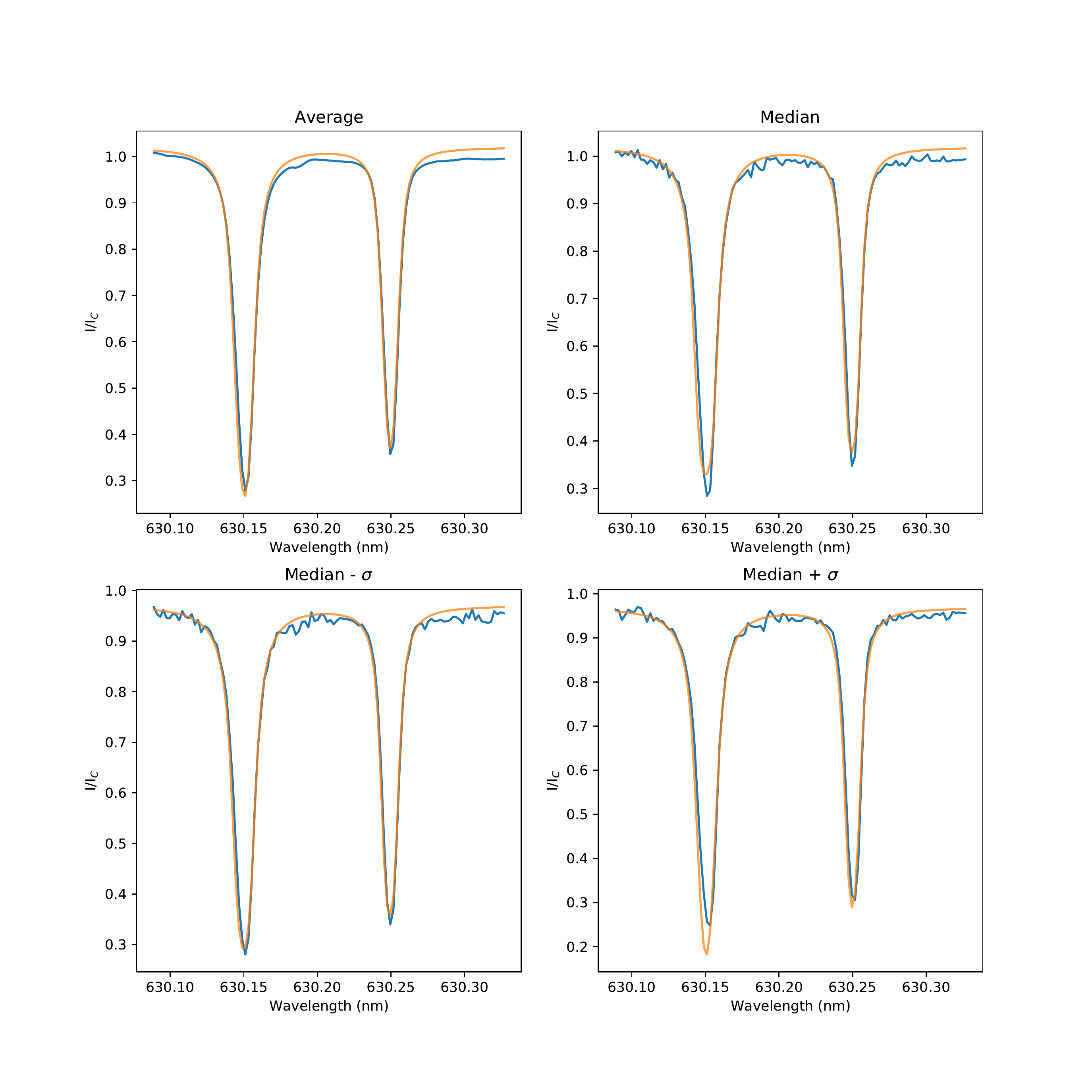}
    \caption{Upper left: Average observed and synthetic profiles in the region. Upper right: Sample profiles where the $\chi^2$ is equal to the median in the region. Lower left: Profiles representative of a good reconstructed fit, where the $\chi^2$ is equal to the median minus one standard deviation. Lower right: Profiles representative of a poor reconstructed fit, where the $\chi^2$ is equal to the median plus one standard deviation. In all cases, the profile in blue is the observation and orange is the synthesis from the reconstructed model atmosphere (reconstructed fit).}
    \label{fits}
\end{figure}

We computed the synthetic profiles from the models reconstructed from the ANN outputs,
obtaining the results shown in Fig~\ref{fits}. These "reconstruction fits" support the
notion that the temperature stratification retrieved by the ANN is consistent with the
observations. The first panel shows the average profiles over the entire region. The other
three are selected representative samples of profiles having a value of $\chi^2$ equal to
the median over the region, and the median plus/minus a standard deviation of all $\chi^2$
values.

\section{Results}

We employed the ANN-based inversions described in the previous sections to explore the
thermal stratification of the solar photosphere. The maps discussed above are in agreement
with previous works in showing a rich thermal structure, rapidly changing with height. In
this paper, we compare the spatial distribution to what is observed in the \ion{Ca}{II} or
G-Band filtergrams.

\subsection{Quiet Sun}

We start by considering the quiet Sun map inverted in the tests of section~\ref{tests}.
The maps are 200$\times$200 pixels but the field of view is not exactly square because the
slit stepping, which establishes the sampling in the $x$-direction, does not necessarily
match the pixel size. In this case the sampling reported in the file headers is
0.15$\times$0.16 arc-seconds per pixel. 

\begin{figure}
    \centering
    \includegraphics[width=.5\textwidth]{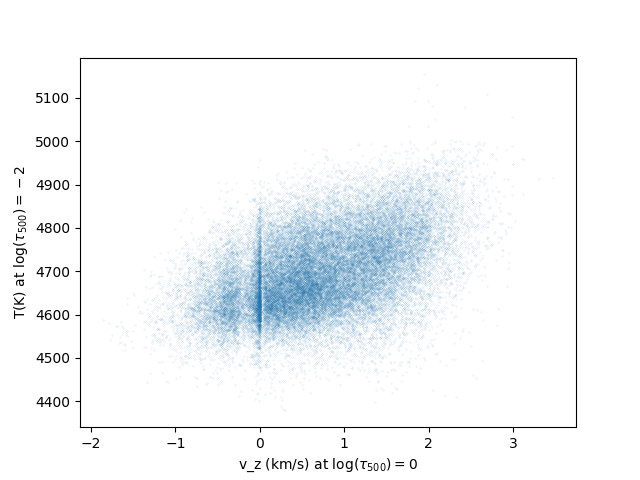}
    \caption{Scatter plot of the mid-photospheric temperatures and the Doppler velocities retrieved by the ANN. }
    \label{fig:rev_gran}
\end{figure}

The first recognizable pattern that stands out is the similarity of the mid-photosphere
temperature map to the Doppler velocity distribution. This is the well known reversed
granulation effect, a natural consequence of convective motions. The tightest correlation
in our dataset, shown in Figure~\ref{fig:rev_gran}, is between temperature at \ltau~$=-2$ and
the Doppler velocity ($v_z$), with a Pearson's correlation coefficient of 0.45 (recall
that, as noted above, the velocities retrieved by our procedure are at the base of the
photosphere). Reversed granulation is characterized also by an anticorrelation with the
\ltau~$=0$ map, which in our data is of $-0.37$. 

The scatter plot exhibits some vertical features. These are the result of the ANN
assigning nearly the same value of $v_z$ to many different profiles instead of smearing
them over the uncertainty range of that parameter. In least-squares inversions, the
solutions for similar profiles tend to be spread over the error bar for that parameter
because each inversion has followed a different path on the $\chi^2$ hypersurface.
However, an ANN might end up assigning a specific value for a parameter (or a narrow range
of values) as a "sticky solution" for a range of input profiles. This means that the
resulting maps will usually be less noisy but the noise level should not be considered an
indication of the uncertainties.

\begin{figure*}
    \centering
    \includegraphics[height=.8\textheight]{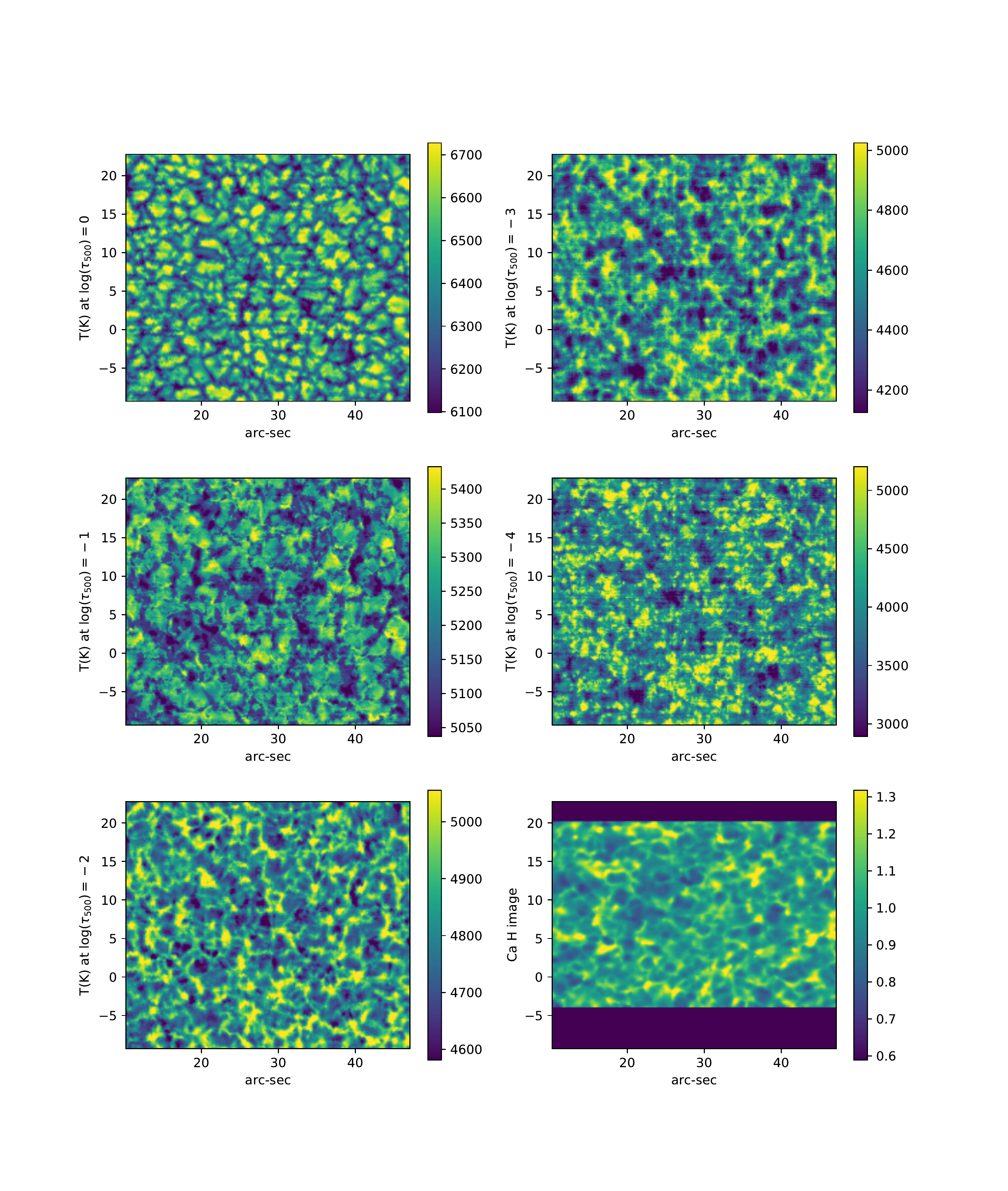}
    \caption{Comparison of temperatures at different heights, as retrieved by the ANN inversion, to \CaII \, filtergram (bottom right panel) in a quiet Sun region.}
    \label{quiet_ca}
\end{figure*}

The temperature maps retrieved in the ANN inversions show a different network structure at
each atmospheric height. Bright photospheric networks have been observed in the wings of
the \ion{Ca}{II} lines and in the G-Band, which are also accessible to Hinode's narrow
band instrument (Hinode/NB). It is then of interest to investigate whether these
structures are related to those, both in the quiet Sun and active regions.

Figure~\ref{quiet_ca} shows a comparison of the temperature maps to a simultaneous
\ion{Ca}{II} image from Hinode/NB. Alignment of SP and NB observations is not
straightforward. We made use of the pointing information stored in the data headers to
bring both datasets to a common reference frame. It should be noted, however, that the
alignment is only accurate to a few arc-seconds {\footnote
{https://hesperia.gsfc.nasa.gov/ssw/hinode/sot/doc/guide/SAGv3.3.pdf}}.

The figure shows no obvious similarities between the \CaII \, intensity and the
photospheric temperatures retrieved at any of the heights. The very weak magnetic fields
in this quiet region do not appear to be correlated to the \CaII \, filtergram, either.

\subsection{Active regions}
\label{sec:ar}

\begin{figure*}
    \centering
    \includegraphics[width=\textwidth]{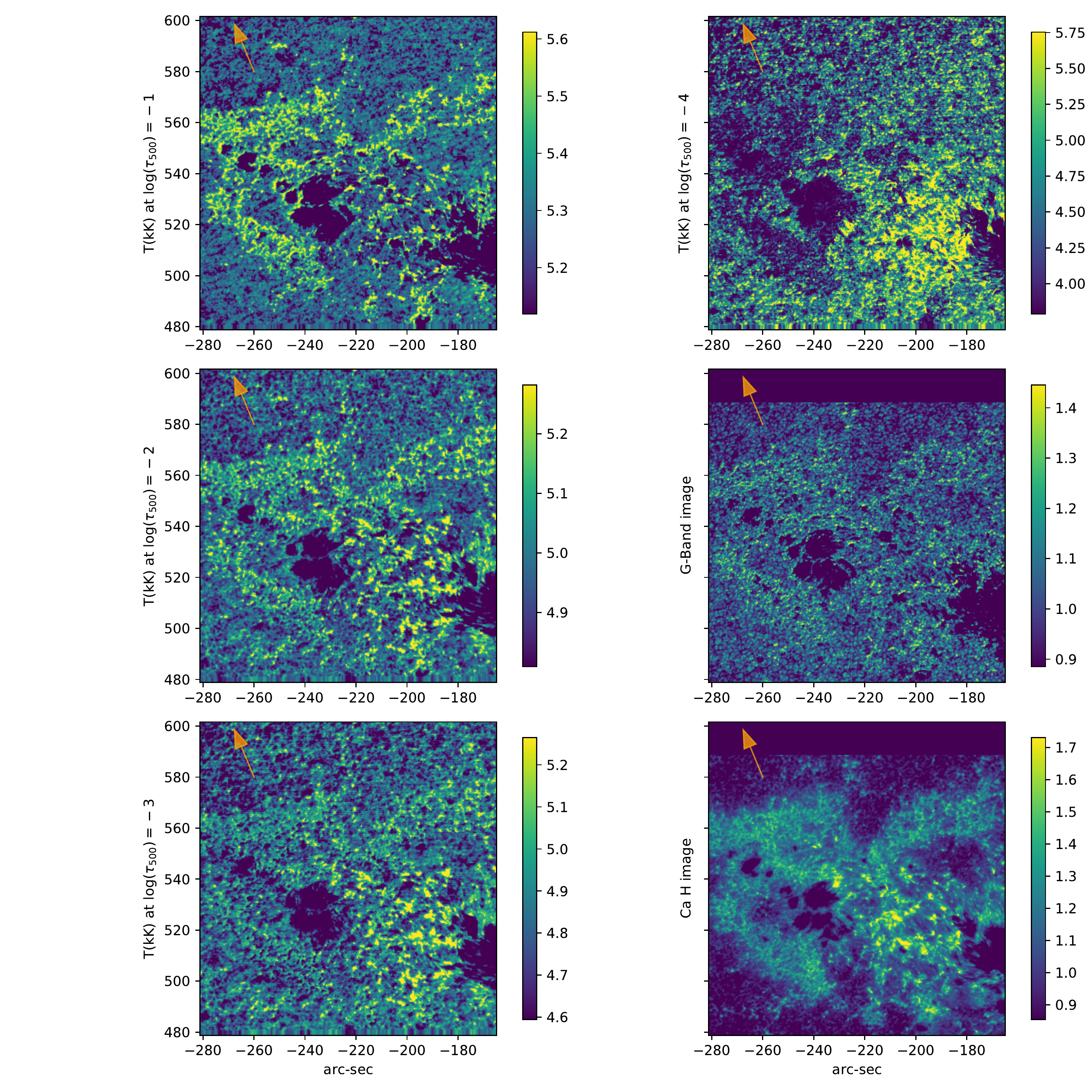}
    \caption{ANN inversion of map1 (active region) at different heights (except \ltau = 0) alongside G-band and \CaII \, filtergrams. The arrow points in the direction to the nearest solar limb.}
    \label{fig:ar1a}
\end{figure*}

\begin{figure*}
    \centering
    \includegraphics[width=\textwidth]{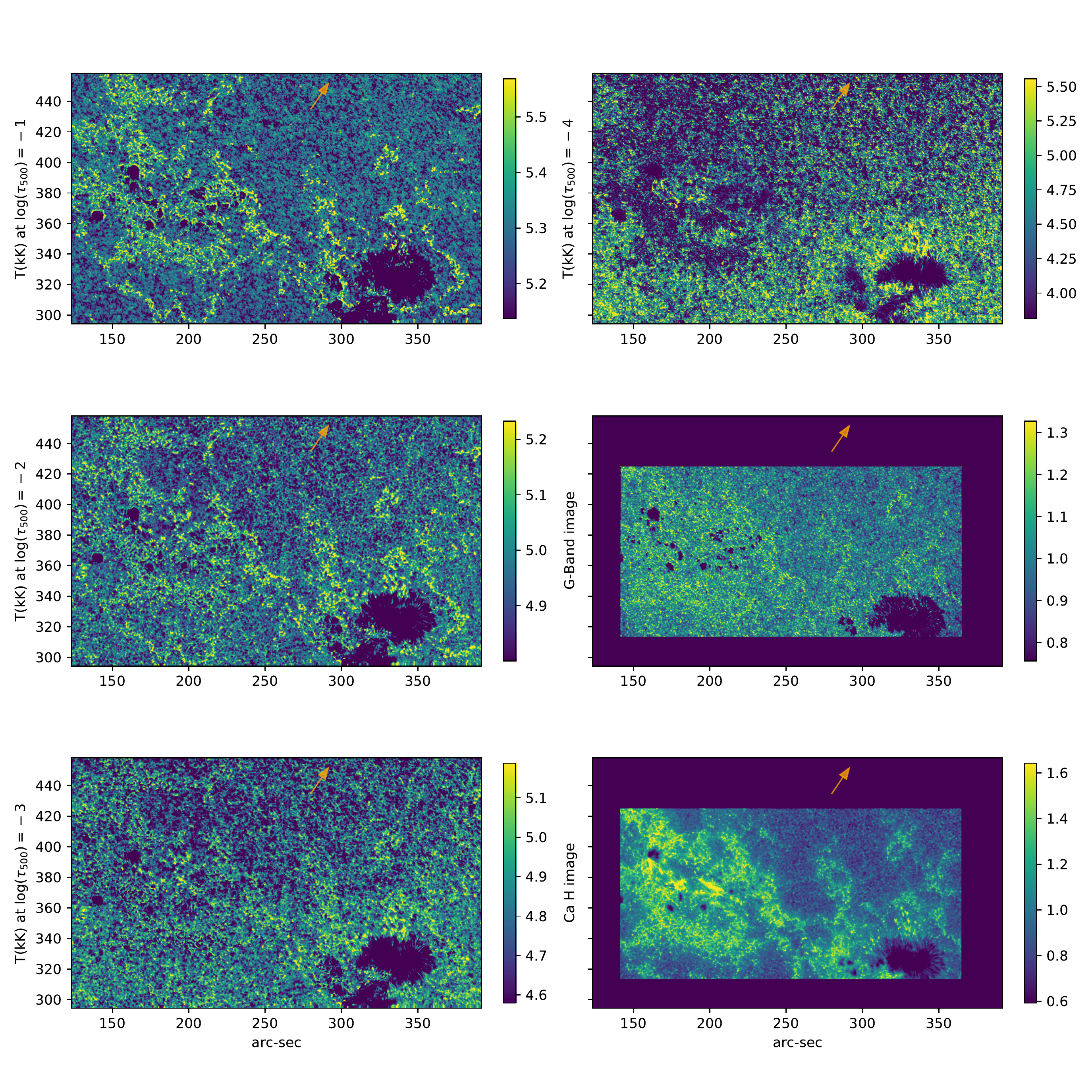}
    \caption{ANN inversion of map2 (active region) at different heights (except \ltau = 0) alongside G-band and \CaII \, filtergrams. The arrow points in the direction to the nearest solar limb.}
    \label{fig:ar2a}
\end{figure*}

In this section we analyze two large-field active-region maps for which there exists
simultaneous G-Band and \CaII \, imaging. The datasets were acquired on January 11, 2010
around 18:30 UT (map1) and January 22, 2012 around 06:30 UT (map2). The spatial sampling
is coarser than in the quiet Sun observations (0.30 and 0.32 arc-seconds in the $x$ and
$y$ directions, respectively) to encompass a larger field of view. Map1 consists of
384$\times$384 spatial pixels, while map2 is 871$\times$512.
The full maps are shown in Figs~\ref{fig:ar1a} to~\ref{fig:ar2a}, with the various panels
displaying temperatures at various heights, along with the narrow-band images (the
magnetic field was already introduced in Figs.~\ref{fig:ar1b} and~\ref{fig:ar2b}). 

The data shows a fine network of hot pixels that roughly follows, in the mid-photosphere,
the magnetic field distribution (\ltau~$=-1$ and $-2$). Higher up the structure is more patchy
and does not follow the  magnetic field maps. Each layer exhibits a different structure
and, more importantly, they also differ from both the G-Band and \CaII \, H images. We
discuss these differences below.

A reversal of hot and cool areas between the middle and the upper photosphere is also
apparent. For instance,  the lower left corner of both maps (Figs.~\ref{fig:ar1a}
and~\ref{fig:ar2a}) is cool at \ltau~$=-1$ but hot at \ltau~$=-4$ (upper left and upper right
panels in both figures). The same anticorrelation is apparent in the upper left and lower
right corners of map2 (Fig.~\ref{fig:ar2a}). In fact, most of the region left of
$x=200$~arc-sec in Fig.~\ref{fig:ar2a} has a reversed appearance. The hot network at
\ltau~$=-1$ (upper left panel) is seen as a dark shadow at \ltau~$=-4$ (upper right panel). The
same is true about the area left of $x=-240$ in map1 (Fig.~\ref{fig:ar1a}). We
quantified this by selecting only the pixels that are hot in either layer ($T>$5400 at
\ltau~$=-1$ or $T>$5000 at \ltau~$=-4$) and computing the Pearson's correlation coefficient. The
values obtained are $-0.77$ in map1 and $-0.78$ in map2.

The temperature maps do not match the narrow-band images. There is some similarity between
the temperature at \ltau~$=-1$ and the G-Band image in the overall distribution of the hot
network. However, a closer look shows important differences (see discussion of
Fig.~\ref{fig:hotwall} below). 

The comparison with the \CaII \, H images is even more puzzling. The Ca emission follows
the pattern of the hot pixels in the mid-photosphere at \ltau~$=-1$ instead of the upper
photosphere, as one would have expected (recall that, as discussed above, the distribution
of hot pixels at \ltau~$=-4$ is anticorrelated with that at \ltau~$=-1$). However, the Ca images
show small-scale filamentary structure in the network, as opposed to the chains of dots
that appear in the temperature maps. The appearance of filaments would suggest that we are
seeing higher layers but the brightness distribution follows that of the mid-photosphere.
We speculate that the most plausible explanation is that the this spectral band has
contributions to its response function from both the low photosphere and the (low)
chromosphere. Detailed radiative transfer modeling would be necessary to confirm this
point but it would require some knowledge of the chromospheric conditions, which is not
available from these data.

Figure~\ref{fig:hotwall} shows a zoom on two regions containing several pores in both
maps. The magnification is different in each figure because the area with pores is larger
in map2. Even though the G-Band bright points extend over the same area as the hot network
at \ltau~$=-1$, they do not exhibit the same features when seen at high resolution.

\begin{figure*}
    \centering
    \includegraphics[width=\textwidth]{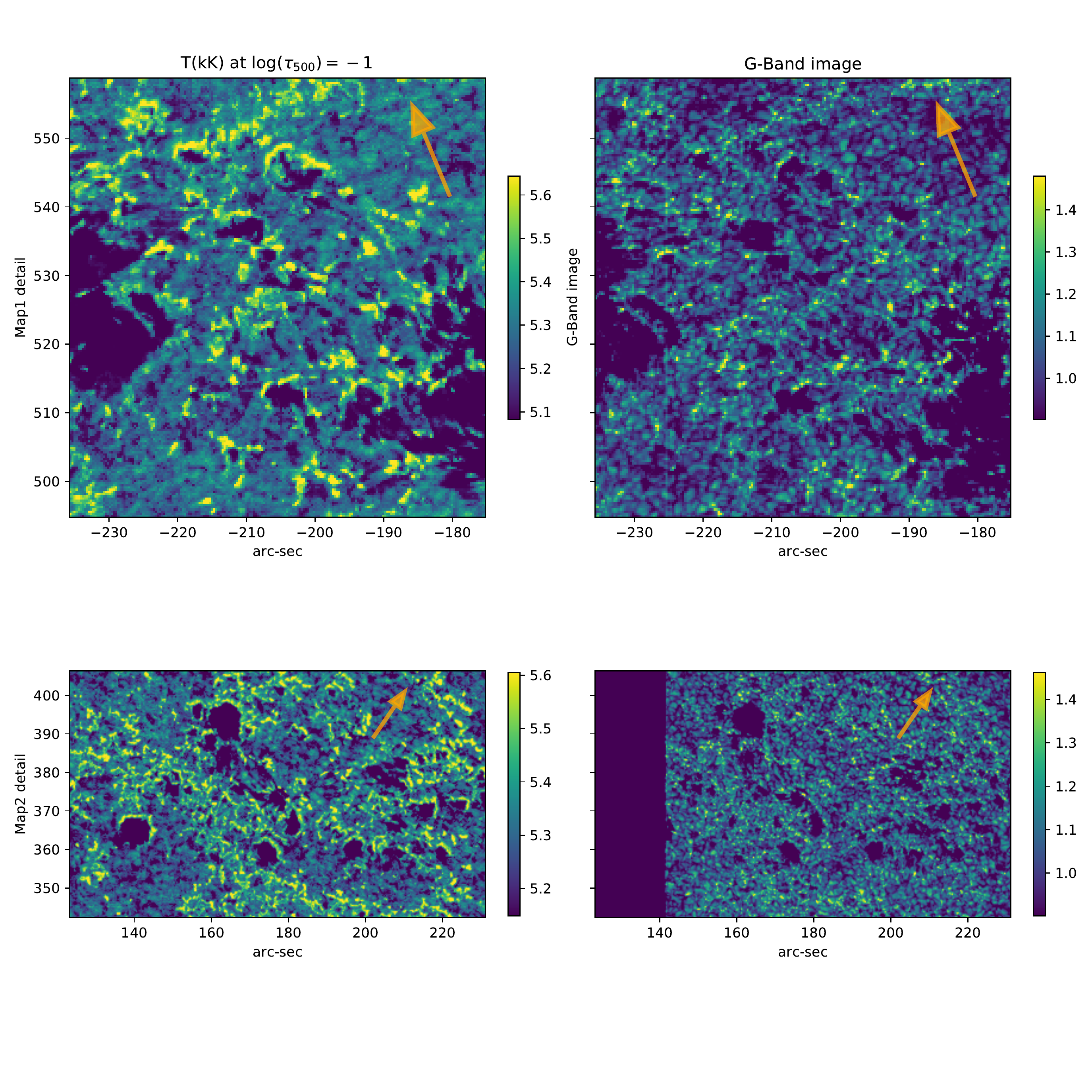}
    \caption{Enlargement of of the areas with pores in the active regions map1 (upper panels) and map2 (lower panels). The left panels show ANN temperature reconstructions at \ltau=-1 and the right panels, the respective G-band filtergrams. Notice the bright arcs around the limb-side (indicated by the orange arrow) of the pores in the left panels.}
    \label{fig:hotwall}
\end{figure*}

A very remarkable feature in these images is the presence of a bright arc around the edge
of pores, tracing the limb side (the arrows indicate the direction to the closest limb).
These arcs are visible around virtually every pore in both datasets. They have a width of
one or two pixels, suggesting that they are not fully resolved in the observations, and
their temperature is always between 5,600 and 5,700~K. By contrast, the pores have
temperatures mostly of 4,600 to 4,800~K but in some cases, particularly the larger ones in
map2, they may go down to 3,600~K, such as in the feature at coordinates (165,390) of
map2.

The bright arcs are probably the pores' "hot walls". The idea of a hot wall seen in
perspective was introduced in early fluxtube models to explain the center-to-limb
variation of faculae and G-band points \citep{S76,KS88,TTT97}. Spruit's original work
considered unresolved fluxtubes and small pores of up to 1,000~km. The pores in our
observations are significantly larger, starting from roughly 3,500~km, but there is no
reason why the same effect should not take place in these. 

\section{Conclusions}

ANN-based inversions are enabling the analysis of large spectroscopic (and
spectropolarimetric) datasets. One such application is presented in this paper. The
training strategy appears to be sufficiently robust for application to real observations
in various situations. 

It is puzzling to find such a clear anticorrelation in the location of hot points in the
middle and upper photosphere. This is counterintuitive and warrants further work to
confirm it, since it appears to challenge the generally accepted idea that small magnetic
elements act as channels to propagate energy into the upper atmosphere (e.g.,
\citealt{JMA+06,RST19}). One possibility is that the energy dissipation and associated
heating might occur at higher layers than we observe here. That would explain the presence
of hot points at intermediate heights that do not exhibit a temperature enhancement in the
upper photosphere. However, this would not explain the patches with hot points in the
upper layers that appear as quiet lower down.

In our ANN approach, each pixel is inverted independently of the rest. Therefore, the
spatial distributions obtained cannot be artifacts of the procedure, they must be present
in the data somehow. A possible mundane explanation for the pixels that are hot in middle
layers and quiet at the top could be that the ANN is not properly trained for such
situations and the closest models in the training set that reproduce the lower and mi
layers are quiet in the upper layers. However, that would not explain the opposite
scenario in the anticorrelation, i.e. the patches with quiet lower and middle photosphere
having enhanced temperature in the upper layers. 

Another important result presented in this paper is the first observation of the "hot
wall" effect, which has been a model prediction since the 1970 (\citealt{S76}) and
explains the bright appearance of small magnetic elements. The original theoretical models
considered smaller pores of up to 1,000~km but we have detected it here in structures of
at least 3,500~km.
Hot walls are believed to be
responsible for the brightness of faculae and small magnetic flux elements
\citep[e.g.,][]{TTT97}. This view, which is now the community consensus, is strongly reinforced by our data.

Our results open the possibility of a future application to chromospheric lines. The
  chromosphere is much more complicated to simulate due to NLTE radiative transfer, much
  faster and vigorous dynamics, faster wave phase velocities, etc. As a result, MHD
  numerical models are not yet sufficiently realistic that they can be used to match the
  observations and this limits the ability to train an ANN with chromospheric simulated
  profiles. However, our training strategy does not require a full numerical model. In
  principle one could apply a similar training using semiempirical 1D models with random
  perturbations to produce a large number of NLTE profiles. The computational effort
  involved in the database generation would be significantly higher than here but still
  feasible.

An area of improvement that we have found with this technique is the "sticky solution",
where the ANN returns basically the same values within a broader uncertainty range,
creating the vertical features seen in the scatter plot of Fig~\ref{fig:rev_gran}.

Finally, we would like to mention a negative result. In spite of our best efforts with
this approach, we have not been able to retrieve gradients in the magnetic field or the
Doppler velocity. We have encountered the same inability to retrieve gradients in previous
works with other ANNs. It is not
clear to us whether this inability is due to a specific problem with our methodology or an
intrinsic limitation of the procedure. An interesting line for future work would be to
explore these and other limitations. 

\begin{acknowledgements}
We acknowledge financial support from the Spanish Ministerio de Ciencia, 
Innovaci\'on y Universidades through project PGC2018-102108-B-I00 and 
FEDER funds. This research has made use of NASA's Astrophysics
Data System Bibliographic Services.  The Python Matplotlib \citep{H07},
Numpy \citep{numpy11}, PyTorch \citep{NEURIPS2019_9015} and IPython \citep{ipython07} 
modules have been employed to generate the figures and calculations in this paper.

\end{acknowledgements}

\bibliographystyle{aa}
\bibliography{aanda.bib,HSN_bib.bib}

\end{document}